\documentclass[runningheads,letterpaper]{llncs}

\usepackage{amssymb,amsmath}
\usepackage{graphicx}
\usepackage[colorlinks]{hyperref}
\usepackage{multirow}

\urldef{\mailsa}\path|ilwoo.lyu@vanderbilt.edu|

\newcommand{\keywords}[1]{\par\addvspace\baselineskip
\noindent\keywordname\enspace\ignorespaces#1}

\makeatletter
\newcommand{\printfnsymbol}[1]{%
  \textsuperscript{\@fnsymbol{#1}}%
}
\makeatother
\begin{document}

\mainmatter  

\title{Cortical Surface Parcellation using Spherical Convolutional Neural Networks}

\titlerunning{Cortical Surface Parcellation using Spherical Convolutional Neural Networks}

%
%
\author{Prasanna Parvathaneni\inst{1}\thanks{P. Parvathaneni and S. Bao contributed equally to this work.}, Shunxing Bao\inst{1}\printfnsymbol{1}, Vishwesh Nath\inst{1},\\Neil D. Woodward\inst{2}, Daniel O. Claassen\inst{3}, Carissa J. Cascio\inst{2}, David H. Zald\inst{4},\\Yuankai Huo\inst{1}, Bennett A. Landman\inst{1}, Ilwoo Lyu\inst{1}}
%
\authorrunning{P. Parvathaneni et al.}

\institute{Electrical Engineering and Computer Science, Vanderbilt University, TN, USA\\
\mailsa\\
\and
Psychiatry \& Behavioral Sciences, Vanderbilt University Medical Center, TN, USA\\
\and
Neurology, Vanderbilt University Medical Center, TN, USA\\
\and
Psychology, Vanderbilt University, TN, USA\\
}

%
%

\maketitle

\setcounter{footnote}{0}

\begin{abstract}
We present cortical surface parcellation using spherical deep convolutional neural networks. Traditional multi-atlas cortical surface parcellation requires inter-subject surface registration using geometric features with high processing time on a single subject (2-3 hours). Moreover, even optimal surface registration does not necessarily produce optimal cortical parcellation as parcel boundaries are not fully matched to the geometric features. In this context, a choice of training features is important for accurate cortical parcellation. To utilize the networks efficiently, we propose cortical parcellation-specific input data from an irregular and complicated structure of cortical surfaces. To this end, we align ground-truth cortical parcel boundaries and use their resulting deformation fields to generate new pairs of deformed geometric features and parcellation maps. To extend the capability of the networks, we then smoothly morph cortical geometric features and parcellation maps using the intermediate deformation fields. We validate our method on 427 adult brains for 49 labels. The experimental results show that our method outperforms traditional multi-atlas and naive spherical U-Net approaches, while achieving full cortical parcellation in less than a minute.
\end{abstract}

\section{Introduction}
Regional-based morphological analysis is a widely adapted approach in neurodevelopmental studies. For valid regional analysis, cortical surfaces need to be consistently subdivided into multi-regions based on cortical parcellation protocols in anatomical or functional fashion \cite{desikan2006automated,klein2010open,yeo2011organization}. However, consistent labeling of cortical regions is challenging due to the complicated cortical folds and inter-subject variability. Typically, manual labeling is tedious and time-consuming, and there exists labeling inconsistency even across experts. In contrast, a multi-atlas cortical parcellation approach \cite{fischl2004automatically} expedites the labeling task with algorithmic consistency. It generally tends to provide better performance as the number of atlases increases. Unfortunately, inter-subject registration is unavoidable in this approach to align multiple atlases to a target subject with significant computational demands proportional to the number of atlases.

With an increasing quantity of imaging data, convolutional neural networks (CNNs) are readily available to handle image segmentation problems on a structured grid. Yet, traditional CNNs architectures are still immature in handling non-uniform data with high complexity. This is mainly because spatial coherence incorporated with existing deep architectures is optimized on standard Euclidean image grids in addition to large memory requirement. In this regard, spherical CNNs recently emerge with efficient operations on a spherical domain. Cohen~\textit{et al.} and Esteves~\textit{et al.} \cite{cohen2018spherical,Esteves_2018_ECCV} proposed spherical CNNs architectures to achieve computational efficiency as well as numerical accuracy. Although they work effectively on classification or regression tasks, semantic segmentation tasks were not fully addressed. Later, general semantic segmentation in a spherical domain was well discussed in \cite{jiang2019spherical}.

Cortical surface mesh is of high complexity that still hampers practical use of existing CNNs due to their limited scalability on large size mesh. A few recent pioneering studies led into drawing the attention of CNNs to surface parcellation, unlike well-developed volumetric segmentation. Cucurull~\textit{et al.} \cite{cucurull2018convolutional} targeted cortical parcellation on only a few ROIs due to memory capacity. Gopinath~\textit{et al.} \cite{gopinath2018graph} proposed better capability with their graph CNNs for full cortical parcellation on adult brains with comparable results to a traditional approach \cite{fischl2004automatically}. The equal importance of training features is also emphasized in recent studies with the central theme being the specific design of the features for accurate cortical parcellation. For example, Gopinath \textit{et al.} \cite{gopinath2018graph} utilized spectral features for better cortical alignments. Wu~\textit{et al.} \cite{wu2018registration} proposed geometry-aware spherical features to use a standard image CNNs architecture.

\begin{figure}[t]
    \centering
    \includegraphics[width=0.92\textwidth]{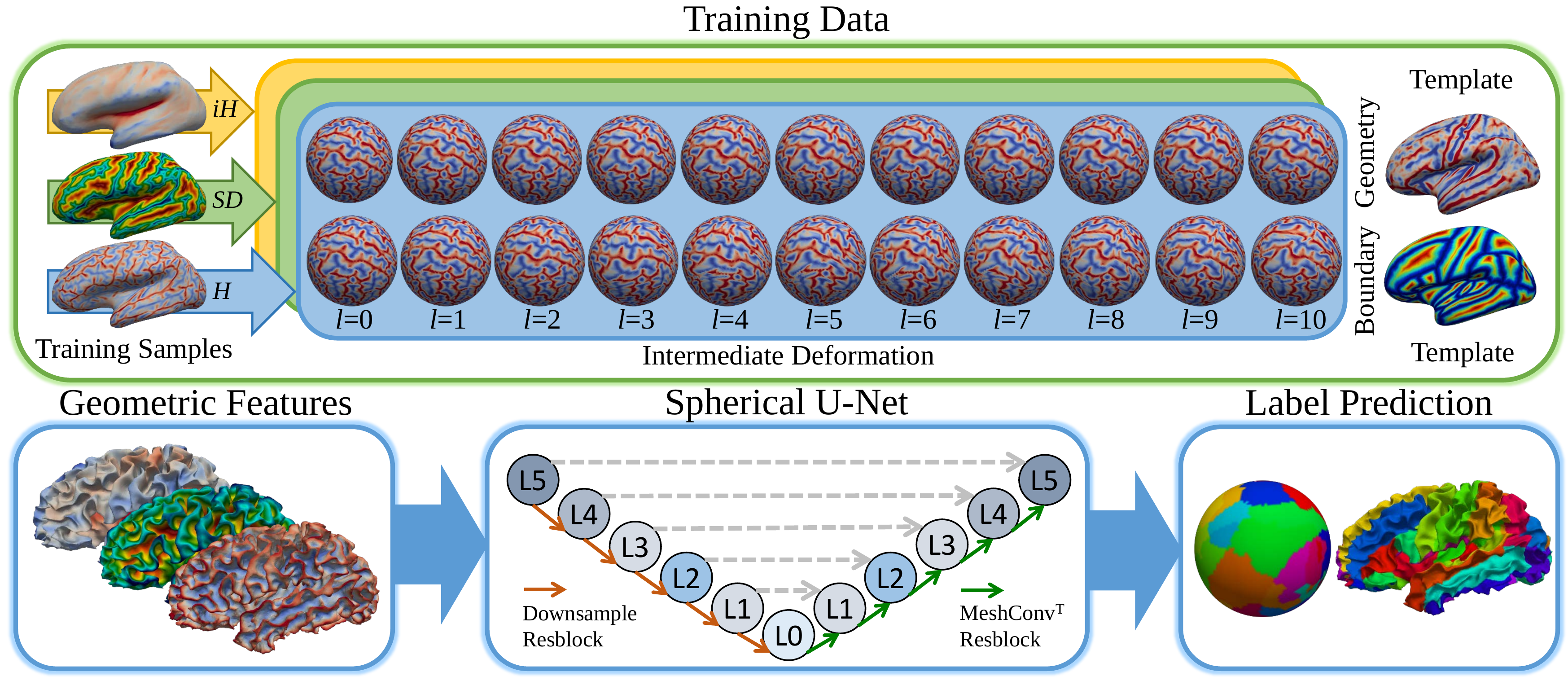}
    \caption{An overview of the proposed method. Three geometric features ($iH$, $SD$, $H$) are used for training the spherical U-Net to predict 49 cortical parcellation labels. For each geometric property, intermediate deformation fields draw a total of 11+11 respective samples after boundary and geometric alignment for data augmentation. The cortical parcellation is then performed using the original geometric features of testing subjects.}
    \label{fig:overview}
\end{figure}

In this paper, we propose a novel cortical parcellation approach using a deep spherical U-Net \cite{jiang2019spherical} that can naturally encode relatively large surface mesh. In particular, we focus on parcellation-specific inputs and their augmentation for efficient utilization of the architecture and accurate parcellation results. Specifically, we compute deformation fields to generate deformed geometric features that best fit ground-truth parcel boundaries using a spherical surface registration method \cite{lyu2019hierarchical}. Since the networks lack generalization of input features, we further propose data augmentation driven by intermediate deformation fields rather than dipole moment variation that overcomes only rotational invariance. This can thus offer a rich set of plausible training samples by leveraging geometric features and their deformation. The key contributions include (i) novel features optimized over cortical parcel boundaries and (ii) data augmentation driven by their intermediate deformation fields. Figure~\ref{fig:overview} shows an overview of the proposed method.

\section{Methods}
\subsection{Objective}
We denote the $i$th cortical label by $z_i \in \mathbb{Z}^+$. Given $N$ cortical labels $\mathbf{L}=\{z_1,\cdots,z_N\}$ and a cortical surface $\Omega \in \mathbb{R}^3$, our objective is to estimate a mapping $F:\Omega \rightarrow \mathbf{L}$ to determine a label for each cortical location.

\subsection{Parcel Boundary Alignment}
\subsubsection{Deformation field.}
Let $M: \mathbb{S}^2 \rightarrow \mathbb{S}^2$ denote a continuous spherical deformation field. Given $\mathbf{x} \in \Omega$ and its corresponding location $\hat{\mathbf{x}}$, the deformation field $M$ holds
\begin{equation}
    \hat{\mathbf{x}}= M(\mathbf{x})\;.
\end{equation}
To estimate $M$, we use a spherical surface registration method \cite{lyu2019hierarchical} that reconstructs $M$ by a linear combination of spherical harmonics coefficients; i.e., $M$ is a function of spherical harmonics degree $l$. For convenience, let $M_l(\cdot)$ denote a deformation field at degree $l$ in the remainder of the paper. An advantage of this method is to easily generate incremental deformation fields by adding basis functions due to orthonormality of spherical harmonics bases that smoothly morph subjects to a target template (e.g., $M_0$ = rigid body alignment, $M_{10}$ = more local non-rigid deformation). Hence, once the deformation fields are computed with a high degree, the intermediate fields can be reconstructed without recomputing low degree again. We can then use all intermediate deformation fields for data augmentation by adding basis functions later.

\subsubsection{Boundary map.}
\begin{figure}[t]
    \centering
    \begin{tabular}{cccc|c}
        \rotatebox{90}{Boundary} &
        \includegraphics[width=0.22\linewidth]{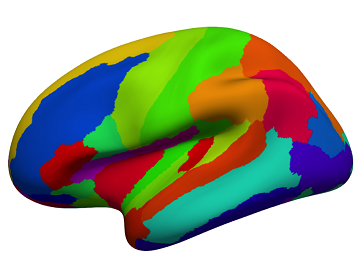}&
        \includegraphics[width=0.22\linewidth]{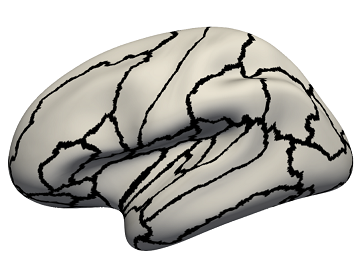}&
        \includegraphics[width=0.22\linewidth]{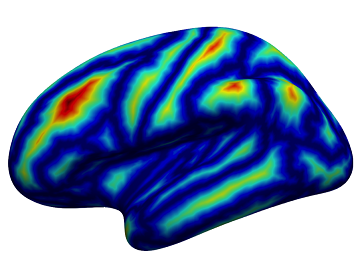}&
        \\
        & Parcellation $F$ & Boundaries $\partial\Omega$ & Distance Map $T$ & \\
        \rotatebox{90}{Alignment} &
        \includegraphics[width=0.22\linewidth]{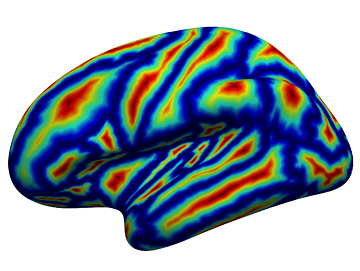}&
        \includegraphics[width=0.22\linewidth]{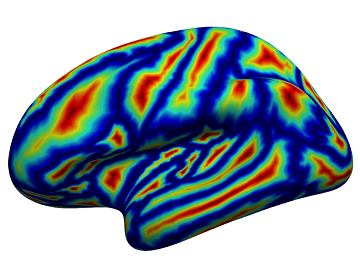}&
        \includegraphics[width=0.22\linewidth]{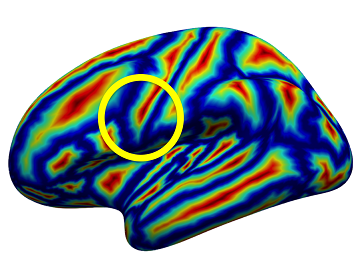}&
        \includegraphics[width=0.22\linewidth]{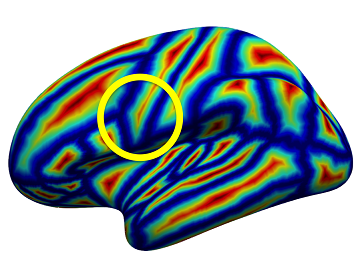}\\
        \rotatebox{90}{$H$ Feature} &
        \includegraphics[width=0.22\linewidth]{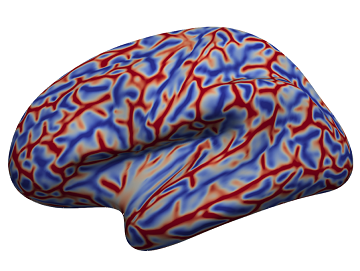}&
        \includegraphics[width=0.22\linewidth]{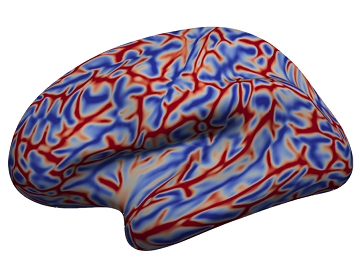}&
        \includegraphics[width=0.22\linewidth]{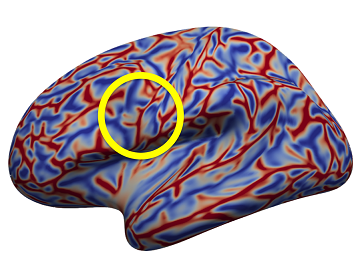}&
        \includegraphics[width=0.22\linewidth]{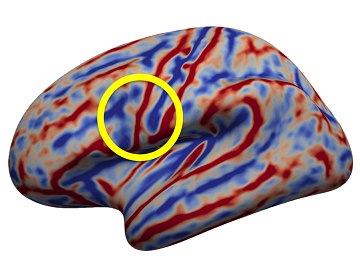}\\
        & $l=0$ (rigid) & $l=5$ & $l=10$ & Template
    \end{tabular}
    \caption{Boundary extraction and alignment. (\textit{1st row}) For inputs for training, parcel boundaries are obtained from ground-truth labels (Eq.~(\ref{eq:boundary})). The boundaries are used to generate distance map $T$ by solving an eikonal equation, and (\textit{2nd row}) smooth trajectory of its deformation to a template is represented by increasing spherical harmonics degree $l$. (\textit{3rd row}) The features for training are accordingly deformed by the deformation fields obtained by the boundary alignment. Note that these boundaries are quite well matched to those of the template, whereas their corresponding deformation on mean curvature $H$ does not fully agree with that of the template (\textit{yellow} circles).}
    \label{fig:boundary}
\end{figure}

Optimal geometric alignment does not necessarily provide optimal cortical parcellation despite their high correlation (see precentral gyrus in Fig.~\ref{fig:boundary} for example). Also, it is important in training to have well-shaped features. Therefore, we compute deformation fields that align parcel boundaries for more accurate prediction. To compute such deformation fields, we need two steps: (1) boundary extraction and (2) the extracted boundaries as a continuous function. Given ground-truth parcel labels $F$, we can obtain boundaries by finding points:
\begin{equation}
    \partial \Omega=\left\{ \mathbf{s} \in \Omega | F(\mathbf{s}) \neq F(\mathbf{x}): \forall \mathbf{x} \in \mathbf{N}(\mathbf{s}) \right\},
    \label{eq:boundary}
\end{equation}
where $\mathbf{N}(\cdot)$ is a set of neighboring vertices on $\Omega$. Now, we need to represent boundaries as a continuous function on $\Omega$ to allow derivatives required for the objective function in \cite{lyu2019hierarchical}. The idea is to compute the geodesic distance between $\partial \Omega$. Let $T(\mathbf{x}): \Omega \rightarrow \mathbb{R}^+$ denote the minimum travel-time $\partial \Omega$ to $\forall\mathbf{x} \in \Omega$. The travel-time $T(\mathbf{x})$ holds the following eikonal equation with a unit propagation speed:
\begin{equation}
\begin{aligned}
& \left\Vert \nabla T(\mathbf{x}) \right\Vert = 1, && \mathbf{x} \in \Omega, \\
& T(\mathbf{x}) = 0, && \mathbf{x} \in \partial \Omega.
\end{aligned}
\label{eq_hamlitonian}
\end{equation}
The solution is thus equivalent to the geodesic distance from $\partial\Omega$. In this work, we use the ordered upwind method \cite{sethian2003ordered}. The distance map $T$ is of different scale for each region across subjects. For better surface registration, we further normalize $T$ with respect to the maximum distance per parcel, similar to the distance map normalization in \cite{lyu2018cortical}, which provides consistent measurements across parcellation maps.

\subsubsection{Deformed data.}
For input features for training, we use standard cortical geometric features: mean curvature ($iH(\mathbf{x})\in\mathbb{R}$) from inflated surface (for global cortical folding agreement), sulcal depth ($SD(\mathbf{x})\in\mathbb{R}$) and mean curvature ($H(\mathbf{x})\in\mathbb{R}$) from cortical surface (for local cortical folding agreement). To create a template, we co-register training samples in an iterative averaging manner \cite{lyttelton2007unbiased}. Here, we compute a distance map of the mode (most frequent) cortical labels across the training set after their registration to the template using the three geometric features. We then register the normalized distance map $T$ to the template distance map at $l=10$, which produces deformation fields $M_{10}$. Note that we found no noticeable improvement of the boundary alignment after $l$ becomes greater than 10 in practice. Finally, the deformed data in training are given by a tuple $P_{10}(\mathbf{x})=\left[iH(M_{10}(\mathbf{x})), SD(M_{10}(\mathbf{x})), H(M_{10}(\mathbf{x})), F(M_{10}(\mathbf{x}))\right]$.

\subsection{Data Augmentation}
The proposed feature deformation is latent. It is valid only if an unseen surface has similar geometric patterns to the deformed features. However, it is unlikely to happen unless a fairly large number of training data are given, which suggests data augmentation to predict unseen data better. Thus, our goal is to generate intermediate deformed features between subjects and the target template. In this way, we can include smooth deformation trajectories as additional plausible training samples. Specifically, we create all intermediate samples as follows:
\begin{equation}
    \bigcup_{l=0}^{10} \left\{\left[iH(M_{l}(\mathbf{x})), SD(M_{l}(\mathbf{x})), H(M_{l}(\mathbf{x})), F(M_{l}(\mathbf{x}))\right]\right\}.
    \label{eq:augmentation}
\end{equation}
To exploit more samples, we also compute deformation fields that align the three geometric features to the template in a similar manner. Figure~\ref{fig:boundary} illustrates an example of deformed features along their deformation trajectory.


\subsection{Deep Learning Architecture}
We adapt a state-of-the-art spherical U-net architecture designed for segmentation tasks \cite{jiang2019spherical} that can be naturally extended to cortical spherical parametrization. In this method, the convolutional kernels are predefined as differential operators for the 1st and 2nd derivatives, which yields fast convolution as well as superior performance over existing spherical networks in their benchmarks. In our framework, three geometric features with their augmentation are provided to input channels and $N$ labels (after the deformation) to output channels. In training, we incrementally reconstruct deformation fields from $0$ to $10$, which generates $11\times2$ times of the original size of the training set (deformation driven by parcel boundary and geometric feature). At the end of the testing stage, we refine predicted parcellation maps with a standard graph cut technique \cite{boykov2004experimental} to remove potential isolated regions and to create smooth parcel boundaries.

\section{Experimental Setup}
We used T$_1$-weighted scans on healthy adults ($n=427$) from 23 to 34 years old, acquired from a Phillips 3T scanner. The cortical surfaces and their spherical mapping were reconstructed via a standard FreeSurfer pipeline with a large number of vertices ($\approx 160k$). We used only left hemispheres. The BrainCOLOR protocol \cite{klein2010open} ($N=49$ ROIs) was used for labeling with manual correction.

We trained the spherical U-Net on NVIDIA Titan Xp with a batch size of 4 at level 5 of the icosahedral subdivision due to memory capacity. We used the cross-entropy loss, and a total of 5,205,008 parameters were optimized by the Adam optimizer. The initial learning rate was set to $0.01$ with a step decay of 0.9 per 20 epochs. We randomly divided our data into three sets: training (80\%), validation (10\%), and testing (10\%). Thus, $385\times11\times2=8,470$ training samples were used in our framework after data augmentation. The optimal weights with the lowest validation loss were chosen up to 100 epochs, and each epoch took about 41 minutes for training of the 8,470 data. For a fair comparison, we applied the same graph-cut technique \cite{boykov2004experimental} on all the baseline methods. To avoid potential errors introduced by misalignment, we also used the aligned features rigidly to the template, i.e., $P_{0}(\mathbf{x})=\left[iH(M_{0}(\mathbf{x})), SD(M_{0}(\mathbf{x})), H(M_{0}(\mathbf{x})), F(M_{0}(\mathbf{x}))\right]$).

\section{Results}
For proof of concept, we trained a spherical U-Net model \cite{jiang2019spherical} with the proposed deformed features driven by only $M_{10}$. From the testing set, we then provided the deformed geometric features $P_{10}$ driven by their optimal boundary alignments. The Dice overlap was $88.53\pm1.05\%$. This indicates that prediction is quite accurate if boundary-driven geometric features $P_{10}$ are provided, which is a strong assumption in practice. We observed low Dice overlap of $78.24\pm4.48\%$ when we fed the rigid features $P_0$ from the same testing set to the networks, which is expected as the networks lack generalization. 
After the proposed data augmentation, we observed Dice overlap of $86.59\pm1.53\%$ closer to that with the deformed features driven optimal boundary alignment.

In comparison, we performed surface parcellation using multi-atlas and spherical U-Net \cite{jiang2019spherical} with $P_0$. In multi-atlas, we propagated labels from all training samples to a single subject after surface registration \cite{lyu2019hierarchical}, and their final labels were determined by majority voting. Such a large number of atlases ($= 385$) generally results in accurate parcellation due to less bias to atlas selection with computational demands (about a day: registration for 3-5 minutes per atlas). Also, the spherical U-Net was trained with $P_0$. We note that the spherical U-Net with $P_0$ is presented in this paper first time for evaluation.

\begin{figure}[t]
    \centering
    \begin{tabular}{c|ccc}
        Ground-truth & Multi-atlas & Spherical U-Net & Boundary Feature\\
        \includegraphics[width=0.24\linewidth,height=0.15\linewidth]{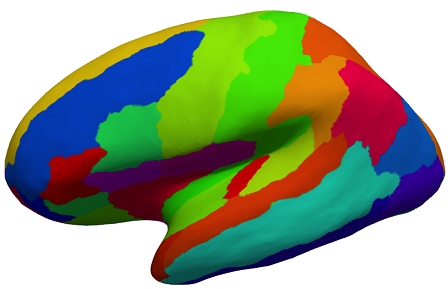}&
        \includegraphics[width=0.24\linewidth,height=0.15\linewidth]{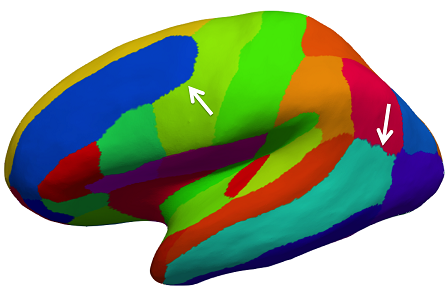}&
        \includegraphics[width=0.24\linewidth,height=0.15\linewidth]{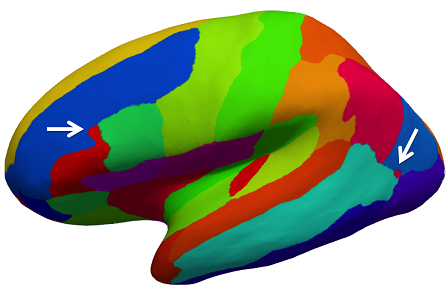}&
        \includegraphics[width=0.24\linewidth,height=0.15\linewidth]{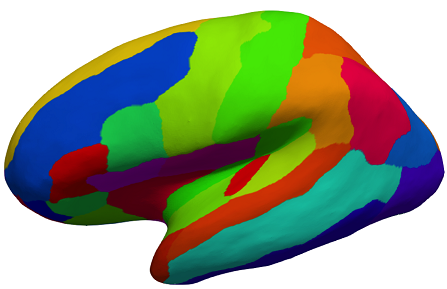}\\
        Mean Dice  & $82.73\pm1.86\%$ & $85.23\pm1.57\%$ & $\mathbf{86.59\pm1.53\%}$\\
        Min/Max Dice & 75.99\%/85.54\% & 80.21\%/88.25\% & \textbf{81.13\%}/\textbf{88.78\%} \\
    \end{tabular}
    \caption{Qualitative comparison: ground-truth, multi-atlas, spherical U-Net, and spherical U-Net with the proposed features. Our approach shows better performance than the other methods. The arrows highlight the mismatching regions to the ground-truth.}
    \label{fig:parc}
\end{figure}

\begin{figure}[t]
    \centering
    \includegraphics[width=1\textwidth]{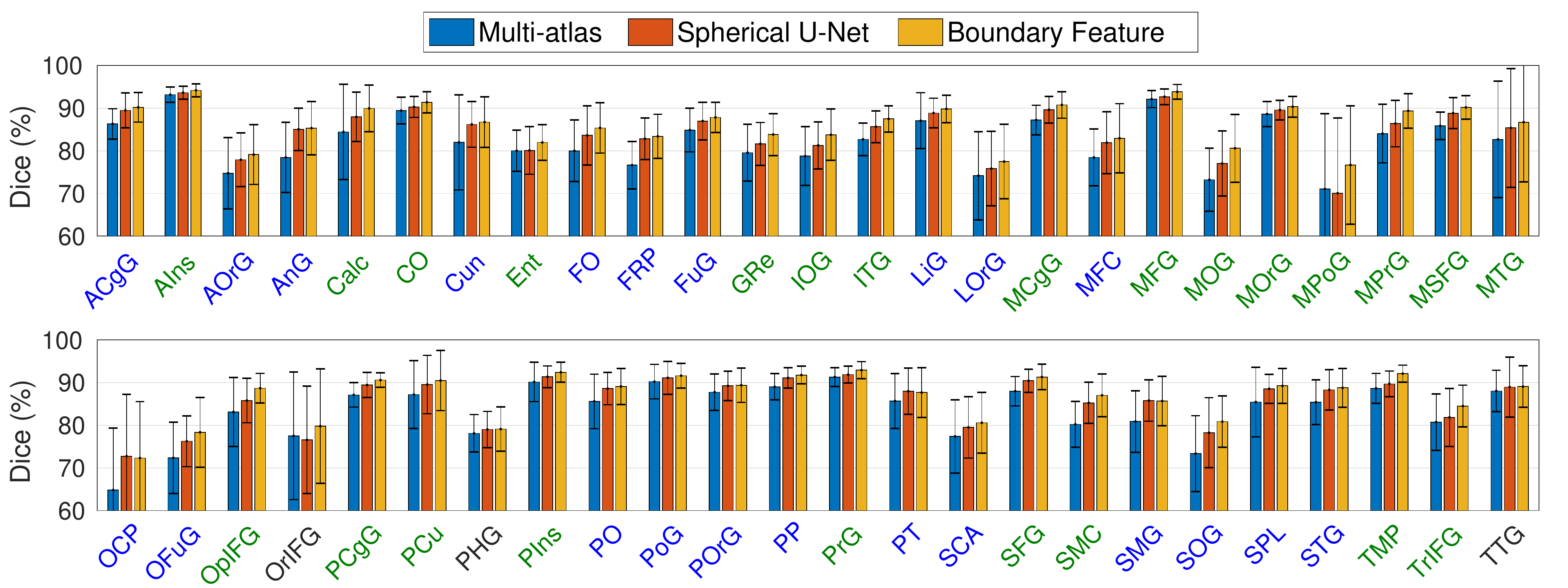}
    \caption{Dice overlap of 49 regions on the left hemisphere. Paired $t$-tests reveal improved regions with statistical significance after the FDR correction ($q=0.05$). 46 and 24 out of 49 regions are improved against multi-atlas and spherical U-Net approaches, respectively. The color in the \textit{x}-axis labels indicates the improved regions: multi-atlas (\textit{blue}), both approaches (\textit{green}), and no improvement (\textit{black}).
    }
    \label{fig:bar_roi}
\end{figure}

The Dice overlap was $82.73\pm1.86\%$ and $85.23\pm1.57\%$ for multi-atlas and spherical U-Net approaches, respectively. Of these approaches, ours achieved the highest Dice overlap with statistical significance in paired $t$-tests ($p<0.05$). Note that both spherical U-Net and our approach used exactly the same input features $P_0$ and no deformed features were provided (i.e., no registration step involved), which hence yields very fast cortical parcellation ($<$ a minute). Figure~\ref{fig:parc} shows an example of resulting cortical parcellation maps for the three approaches. We further performed paired $t$-tests to observe ROI-wise improvement on individual parcels. The test statistics revealed that our approach significantly improved parcellation accuracy after false discovery rate (FDR) \cite{benjamini1995controlling} for multi-comparison correction ($q=0.05$). Our approach outperforms multi-atlas (46 regions) and spherical U-Net (24 regions) as shown in Fig.~\ref{fig:bar_roi}. It is noteworthy that no regions were found with significantly reduced Dice overlap.

\section{Conclusion}
We presented a cortical parcellation method using spherical U-Net with novel features optimized over cortical parcellation boundaries. To enhance the capability of the spherical U-Net, we also incorporated intermediate deformed features along trajectories of the deformation fields. In the experiments, the proposed method achieved qualitatively and quantitatively better performance. Furthermore, full cortical parcellation was obtained in less than a minute.

\subsubsection*{Acknowledgments.}
This work was supported in part by the National Institutes of Health under Grants R01EB017230, R01MH102266, R01NS097783, R01MH102272, and R01MH098098, in part by the National Science Foundation under Grant CAREER IIS 1452485, in part by the VISE/VICTR under Grant VR3029, and in part by NVIDIA Corporation under GPU Grant Program.

\bibliographystyle{splncs04}
\bibliography{reference}

\end{document}